# PrivyTRAC – Privacy and Security Preserving Contact Tracing System

Ssu-Hsin Yu[1]

[1] Scientific Systems Company, Inc., 500 W. Cummings Park, #3000, Woburn, MA 01801, USA
`syu@ssci.com`

**Abstract.** Smartphone location-based methods have been proposed and implemented as an effective alternative to traditional labor intensive contact tracing methods. However, there are serious privacy and security concerns that may impede wide-spread adoption in many societies. Furthermore, these methods rely solely on proximity to patients, based on Bluetooth or GPS signal, ignoring lingering effects of virus, including COVID-19, present in the environment. This results in inaccurate risk assessment and incomplete contact tracing. A new system concept called PrivyTRAC preserves user privacy, increases security and improves accuracy of smartphone contact tracing. PrivyTRAC enhances users' and patients' privacy by letting users conduct self-evaluation based on the risk maps download to their smartphones. No user information is transmitted to external locations or devices, and no personally identifiable patient information is embedded in the risk maps as they are processed anonymized and aggregated confirmed patient locations. The risk maps consider both spatial proximity and temporal effects to improve the accuracy of the infection risk estimation. Experiments conducted in the paper illustrate improvement of PrivyTRAC over proximity-based methods in terms of true and false positives. An approach to further improve infection risk estimation by incorporating both positive and negative local test results from contacts of confirmed cases is also described.

## 1 Introduction

As severe travel restrictions due to COVID-19 are being gradually relaxed in order to minimize further damage to the economy, it is expected that there will be continued cases of infection and pockets of community transmission until vaccines become widely available. To prevent sporadic cases from becoming sources of another outbreak, rigorous contact tracing is essential. Compared to the traditional approach of interviewing patients, the smartphone location-based approach has proven to be an effective alternative to accomplish comprehensive contact tracing with much less labor demands. However, there are serious privacy and security concerns associated with the smartphone-based methods that may impede adoption and wide-spread use in the many societies [1]. Effectiveness of the smartphone contact tracing approach can be significantly improved with active cooperation from the public by alleviating the security and privacy concerns.

We propose a privacy-preserving, secure and accurate COVID-19 contact tracing systems called PrivyTRAC: Privacy and Security Preserving Contact Tracing System. PrivyTRAC

1. is a smartphone location-based contact tracing system with security and privacy inherent in the tracing mechanism that protects the privacy of both the patients and the public, and
2. infers from confirmed cases COVID-19 infection risk at given places and time that allows individuals to self-assess exposure risk from their movement histories.

The proposed system, illustrated in Fig. 1, is built on two innovations (see Section 3 for details). One is the innovative mechanism that utilizes the infection risk maps (Step 1 in Fig. 1), which are processed from anonymized, aggregated patient locations information, that individual users can download to a smartphone App (Step 2). The users can then evaluate locally their own risks of contracting COVID-19 due to contacts (Step 3). This mechanism does not require users to upload their personally identifiable information (PII) to an external platform, nor do the users need to broadcast their PII to other smartphones in vicinity.

The second innovation is the estimation of the spatio-temporal infection risk maps that allow users to self-assess their own risks. The risk maps consider not only the spatial proximity to COVID-19 patients, as in most contact tracing approaches, but also the temporal effect of how long the virus can stay virulent in the environment. We have learned that the novel coronavirus can remain contagious on surface for an extended period of time [2,3] and virus-containing droplets can travel significant distances [4,5]. The risk-based approach that considers not only instantaneous proximity to the patients but also the lagging effects after a patient has left thus offers a more accurate risk assessment of contracting the virus. Furthermore, as researchers continue to learn the factors that influence the virus spread, new findings can be quickly incorporated into our risk estimation model to refine the risk maps.

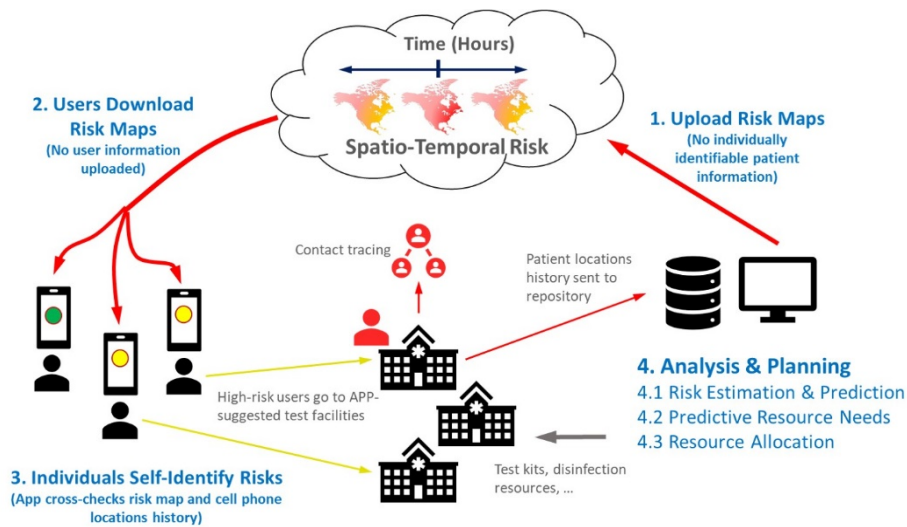

**Fig. 1.** PrivyTRAC Contract Tracing Process and Information Flow

## 2   Current Smartphone Location-based Contact Tracing Methods

Many smartphone location-based contact tracing approaches have been proposed or implemented recently [2,5,6,7,8,9,10]. They generally fall into two categories. One is to aggregate all smartphone movements of a population, whether a person is ill or not, in a centralized repository to identify encounters with the confirmed COVID-19 cases. The other category of methods is to use a phone's Bluetooth radio to record encounters of all other phones in proximity [3] and later alert the user when the unique cell phone ID of a confirmed case matches an ID in the user's phone.

These two approaches suffer two main drawbacks – the first is privacy and security, and the second is accuracy. Their mechanisms require exposing individuals' locations and unique IDs, either in a centralized, externally maintained repository or to all other phones in vicinity. Privacy concerns may cause people to be less willing to adopt the tools, and hence render the tools less effective. Moreover, whether individuals' movement data and their encounters with other people are stored centrally or locally, there is always inherent risk that sophisticated and determined hackers can exploit software weaknesses to acquire personally identifiable information (PII) of individuals and their encounters, similar to the actions taken by some data brokers and aggregators for advertising purposes. In fact, Cybersecurity and Infrastructure Security Agency (CISA) has issued warning of Advanced Persistent Threat (APT) actors exploiting the COVID-19 pandemic to collect bulk personal information [4].

Besides privacy and security concerns, those contact-tracing tools offer an incomplete measure for determining the infection risk. As we have learned, novel coronavirus can survive on surface for an extended period of time and its droplets can travel for extended distances, depending on the surface materials and other environmental factors. A person does not need to be in the immediate vicinity of a patient to be exposed to the virus through indirect contacts. Hence, determining a person's risk of contracting the virus based solely on direct proximity to the infected does not provide an accurate risk assessment.

## 3     PrivyTRAC Approach

### 3.1     System Architecture

Our contact-tracing approach PrivyTRAC is illustrated in Fig. 1. The system consists of two main components – one residing on individual users' smartphones and the other on the server. On the user side, the smartphone App regularly (or as requested by the users) downloads up-to-date spatio-temporal exposure risk maps from the server. The risk maps quantify the likelihoods of infection at particular locations and time. Based on the maps, the App would then cross-check with the user's smartphone locations data. Using the risk maps and the locations data, the App computes the aggregated risk of contracting the virus. If the person's risk exceeds a certain threshold, the App would notify the person and suggest follow-on actions, such as the testing sites for confirmation.

Under this process, a user's private locations history and PII never leave his/her own phone nor are they being recorded by other phones. Patients' privacy is preserved as well, since the patients remain anonymous and the downloaded risk maps contain only processed and aggregated locations information from many patients, making it extremely challenging to extract PII. Furthermore, the ability for users to maintain control of their private data and decide when the service is activated will significantly encourage adoption by the public.

On the server side, the spatio-temporal risk maps are computed and continually updated as new cases are reported. The movement histories of confirmed cases are collected by public health agencies. The sever software computes the infection risks at particular locations and time, based on the aggregated locations of the patients from the public health agencies and factors affecting the virus' persistent virulence. Those factors include distance from an infected person, durations of the virus' survival on various surface materials, length of exposure, and environmental conditions such as temperature and sunlight. By incorporating the disease vector and environment's effects on virulence rather than merely considering direct encounters based on smartphone proximity, the resulting infection risk estimation is more comprehensive and accurate.

### 3.2     Infection Risk Map Computation

To evaluate the risk of contracting the disease from an infected person, we assume that the probability of infection decreases exponentially with distance and time. For the risk analysis, we consider a spatio-temporal grid consists of 1 meter by 1 meter by 1 second cells. If a person with the disease stays in a 1 meter by 1 meter area centered at location $(x_p, y_p)$ for 1 second around time $t_p$, a person in an area of the same size around location $(x, y)$ for 1 second around time $t$ is assumed to have the probability of contracting the disease ($C = 1$) as follows:

$$P(C = 1|x, y, t, x_p, y_p, t_p) = \begin{cases} p_0 \exp\left(-\frac{(x-x_p)^2}{\sigma_x^2} - \frac{(y-y_p)^2}{\sigma_y^2} - \frac{(t-t_p)^2}{\sigma_t^2}\right) & \text{if } t \geq t_p \\ 0 & \text{otherwise} \end{cases} \quad (1)$$

If there are a total of $N$ such spatio-temporal cells $(x_p^i, y_p^i, t_p^i)$ with a patient present, regardless of whether they are occupied by the same patient or not, a person in the 1 m. by 1 m. by 1 sec. cell $(x, y, t)$ has the probability of being infected as follows:

$$P(C = 1|x, y, t, x_p^i, y_p^i, t_p^i, \ i = 1, \ldots, N) =$$
$$1 - \prod_{i=1}^{N}\left(1 - P(C = 1|x, y, t, x_p^i, y_p^i, t_p^i)\right) \quad (2)$$

Note that the cells $(x_p^i, y_p^i, t_p^i)$ are not necessarily due to the same patient. Hence, we can easily aggregate multiple patients in the same risk map.

If a person is in the area for a certain duration, we consider each contact within one second as an independent event. Hence, the person's overall probability of contracting the disease is

$$P(C = 1| s) = 1 - \prod_{j=1}^{M} \prod_{i=1}^{N}\left(1 - P(C = 1|x^j, y^j, t^j, x_p^i, y_p^i, t_p^i)\right) \quad (3)$$

where $s$ represents a sequence of $M$ such 1 m by 1 m by 1 sec spatial-temporal cells $(x^j, y^j, t^j)$ that the person of interest occupies in the vicinity of infected people.

Based on the above analysis, an area risk map can be created from Eq. (2). By using aggregated locations and time of people with the disease $(x_p^i, y_p^i, t_p^i)$, the risk of contracting the disease $P(C = 1|x, y, t, x_p^i, y_p^i, t_p^i, \ i = 1, \ldots, N)$ at any location and time $(x, y, t)$ can be computed. When a user downloads the spatio-temporal risk map into their smartphone App, their own individual risk $P(C = 1| s)$ can be computed by the App according to the location history $s$ on the smartphone using Eq. (3). The App can apply a risk-based metric, for example, to advise whether a person should seek further medical help based on their probability of contracting COVID-19 due to the contacts.

To illustrate the varying risk of contracting the disease depending on the distance from an infected person and the elapsed time since the person's presence, we plot in Fig. 2 the log-probability of infection. For this example, it is assumed that a patient is present at time 0 at location $x_p = 0$ and $y_p = 50$ for 1 second. We also assign $p_0 = 0.01/\sqrt{2\pi}$, $\sigma_x = \sigma_y = 1$ and $\sigma_t = 100$ in Eq. (1). The resultant log-probabilities $P(C = 1|x = 0, y, t, x_p = 0, y_p = 50, t_p = 0)$ of infection risk for different $y$ locations (vertical axis) at different time (horizontal axis) after the 1-second presence of the patient at time 0 are shown in Fig. 2. The figure shows that despite the patient not being present (or in proximity) after the initial 1 second, the risk of contracting the disease is still present, albeit small as elapsed time increases. Hence, to fully account for the risk of contracting the disease for contact tracing purposes, it's essential not only to consider the immediate proximity to the patient but also the lingering effects of the virus in time.

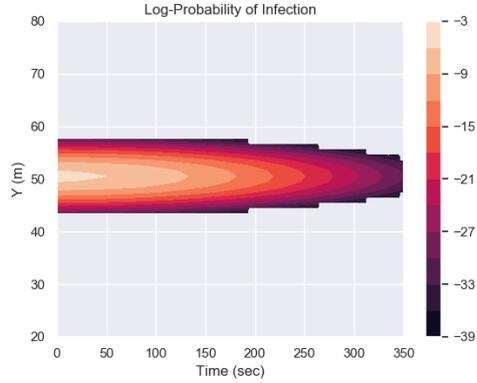

**Fig. 2.** Log-probability of infection risk when a patient is at location $y = 50$ for 1 second

## 4 Simulation Experiment Results

To illustrate the difference between the risk-based and the proximity-based contact tracing approaches, we conducted simulated experiments of the two methods. In the experiments, we consider an area of 100 meters by 100 meters (Fig. 3) in a 350-second span. At the beginning of this time span $t = 0$, a patient enters the square area from the middle of a side and travels across the area parallel to another side at the constant speed of 1 m/sec, as shown in red in Fig. 3. In the time span, a person enters the same area at random time, location and speed. The time was chosen with a uniform distribution between time 0 and 200 seconds; the location was chosen with a uniform distribution from the middle halves of the 4 sides enclosing the area; the speed remains constant as the person traverses the area and was chosen with a uniform distribution between 0.75 m/sec and 1.25 m/sec. Under the above simulation conditions, the experiments were repeated 20,000 times. Fig. 3 shows the trajectory of the patient in red in (a) spatial and (b) spatial and temporal coordinates, as well as the trajectories of 4 healthy individuals in blue.

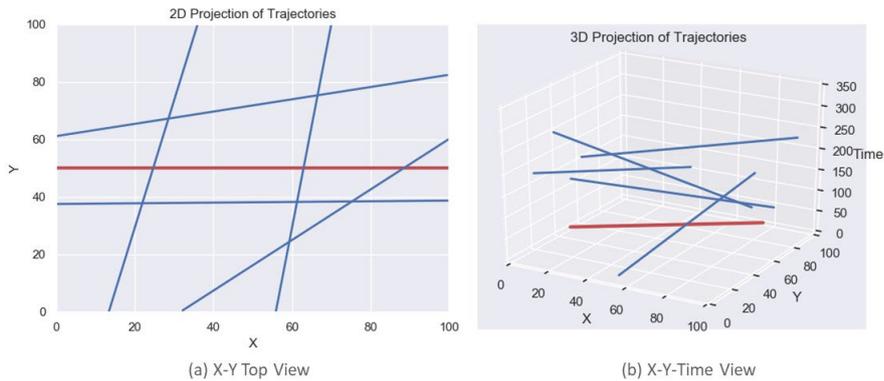

(a) X-Y Top View      (b) X-Y-Time View

**Fig. 3.** Trajectory of a patient (red) and 5 sample trajectories of healthy individuals (blue)

Using Eq. (2), we can compute the infection risk map in the square area induced by the presence of the patient. For each healthy individual that traverses the area, Eq. (3) provides the person's risk of contracting the disease. A person is advised to seek further testing if their risk exceeds a certain threshold.

For comparison purposes, we also implemented a spatial proximity-based metric. That is, a person is advised to seek testing if the person is within a certain distance from the patient at any time.

We plot in Fig. 4 the probability of correctly identifying a person that actually contracted the disease (true positive) versus the probability of falsely advising a healthy person to seek medical help (false positive) by varying the risk threshold for the risk-based approach (blue lines) and the distance threshold for the proximity-based approach (red lines). An ideal system would have true positive probably 1 at 0 false positive probability, i.e. the upper left corner in the plots. The 4 plots in Fig. 4 from left to right show the different choices in Eq. (1) for $\sigma_t = 10, 50, 100, 150$ seconds respectively. If the ability for the virus to infect diminishes quickly over time (e.g. $\sigma_t = 10$ sec), the difference between risk-based and proximity-based contact tracing is small Fig. 4(a). On the other hand, as the decay time increases ($\sigma_t = 50, 100$ sec), the risk-based approach performs significantly better than the proximity-based approach Fig. 4(b)(c). It becomes apparent in the cases where the virus' ability to infect diminishes slowly that the proximity-based approach cannot fully identify the infected people without incurring unacceptable false positives.

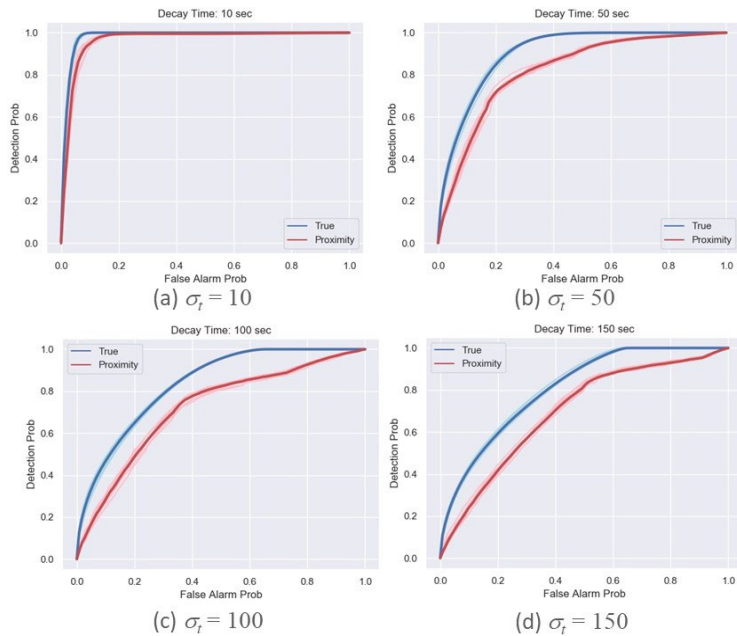

**Fig. 4.** True positive vs. false positive for different $\sigma_t$ decay rates; lighter-colored lines are from results using 1/10 of the simulated trajectories.

The reason the risk-based and the proximity-based measures differ can be seen in Fig. 5. In the figure, we plot the probability of contracting the disease on the x axis and, on the y axis, the 1 over the exponential of minimum distance from the infected person (i.e. $e^{-\min \text{distance}}$), for $\sigma_t = 50$ in Eq. (1). $e^{-\min \text{distance}}$ is chosen such that a higher number means a person is closer to the patient at some point in time and hence subjects to a higher probability of being infected. Each point on the scatter plot represents a case in the experiment. The marginal distributions (histograms) of the cases are also plotted on the top and on the right for the risk-based and the proximity-based measures respectively. If the proximity- and risk-based measures are similar, we would expect to see positive correlation on the plot. Even though most cases that are in proximity to the patient tend to have higher probabilities of infection, there are significant number of cases that are far enough from the patient spatially but are still subjected to high infection probabilities. It is mainly due to the delayed effect when the patient has left a location but the virus in the environment still has the ability to infect.

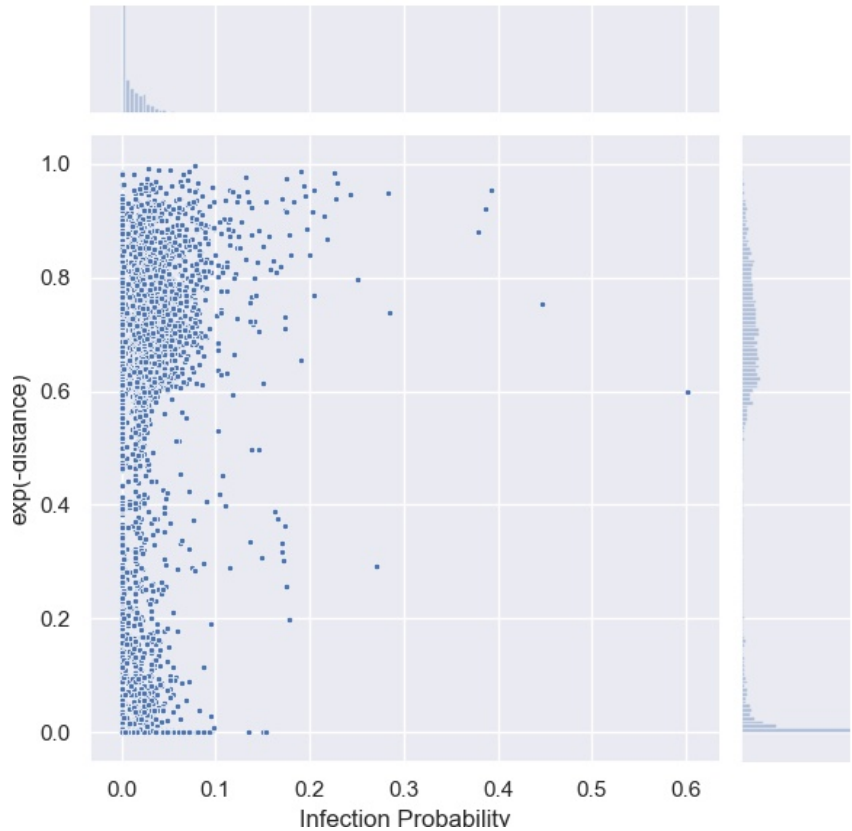

**Fig. 5.** Joint distribution of log-probability of infection and exp(-distance)

## 5  Model Refinement

The experiments conducted in this paper utilize the probabilistic infection risk model in Eq. (1) that assumes exponential decay of infection risk in space and time. Since various environmental factors can impact how far virus-containing droplets can reach and how long the virus can remain contagious in air and on surface, the models can be extended to improve its accuracy by incorporating those factors if they are available. For example, the variances $\sigma_x^2$ and $\sigma_y^2$ in Eq. (1) can be replaced by a 2-dimensional covariance matrix to capture the effects of prevailing wind speed and direction on the spread of the droplets. The temporal variance $\sigma_t^2$ can be a function of the local temperature and humidity. In a confined environment, where infection by way of indirect contacts is possible, we can incorporate surface properties as part of the model.

Another direction to refine the infection risk estimation is to consider reported cases of infection as observations. Consider the scenario where several people that came into contacts with a patient in a particular area have been tested for infection. Some of them were tested positive while others negative. The test results and their movement histories in the area are observations of the underlying risk model. The initial risk model is constructed mainly based on the knowledge of average reach and decay time of the virus. The new test results, positive or negative, provide additional information to adjust the model parameters so as to better align with local conditions. A Bayesian probabilistic model is well suited for this purpose:

$$\tau \sim \text{Gamma}(\alpha, \beta), \quad \tau_t \sim \text{Gamma}(\alpha_t, \beta_t) \tag{4}$$

$$P(C = 1 | x, y, t, x_p^i, y_p^i, t_p^i) = \begin{cases} p_0 \exp\left(-\tau(x - x_p^i)^2 - \tau(y - y_p^i)^2 - \tau_t(t - t_p^i)^2\right), & \text{if } t \geq t_p^i \\ 0 & \text{otherwise} \end{cases} \tag{5}$$

$$P(C = 1 | s) = 1 - \prod_{j=1}^{M} \prod_{i=1}^{N} \left(1 - P(C = 1 | x^j, y^j, t^j, x_p^i, y_p^i, t_p^i)\right) \tag{6}$$

$$T \sim \text{Bernoulli}(P(C = 1 | s)) \tag{7}$$

Eq. (5) is similar to Eq. (1) except that the variances are re-defined as precisions for the convenience of specifying their prior distributions in Eq. (4). The prior distributions of $\tau$ and $\tau_t$ are Gamma distributions of hyper-parameters $(\alpha, \beta)$ and $(\alpha_t, \beta_t)$ respectively, where the symbol tilde (~) denotes "distributed as." Eq. (6) is the same as Eq. (3). Recall that $s$ represents a sequence of $M$ such 1 m by 1 m by 1 sec spatial-temporal cells $(x^j, y^j, t^j)$ that a person of interest occupies in the vicinity of the patients, and $P(C = 1 | s)$ is the probability of the person contracting the disease. The result of the test, positive ($T = 1$) or negative ($T = 0$), is modeled as the outcome of a Bernoulli trial as in Eq. (7) where the parameter of the Bernoulli distribution is the infection probability $P(C = 1 | s)$.

Based on the above Bayesian probabilistic model, we can refine the model parameters $\tau$ and $\tau_t$ by computing their posterior distributions:

$$P(\tau, \tau_t \mid T_1, T_2, \ldots, T_L) \tag{8}$$

where $T_1, T_2, \ldots, T_L$ are $L$ individuals' test outcomes, which are either positive $T_i = 1$ or negative $T_i = 0$. Computational methods such as Markov chain Monte Carlo (MCMC) [12] can be applied to compute the posterior distribution.

The refined model parameters $\tau$ and $\tau_t$ and the resultant spatio-temporal infection risk probability in Eq. (5) then provide a more precise local risk map. By using the improved risk map, we can further reduce the chance of missing positive cases and optimize the use of resources by avoiding unnecessary testing.

## 6 Conclusions

The impacts of severe travel restrictions are enormous and their cost has been felt throughout the economy. Effective contact tracing is a key step in relaxing those measures while keeping the infection under control. Traditional contact tracing measures based on interviews with patients are labor intensive and error-prone. Contact tracing through personal electronic devices such as smartphones has been proposed as an effective measure to overcome these challenges. Although smartphone-based contact tracing has been successfully implemented in some countries, the privacy implication and security concerns can impede broad adoption of similar measures in other societies. Without wide-spread adoption, the effectiveness of electronic contact tracing can be severely limited.

The proposed electronic contact-tracing approach PrivyTRAC respects privacy of individuals and increases security of the system through a de-centralized mechanism. By preserving privacy and enhancing security, it will significantly promote cooperation from the public and facilitate adoption by public health agencies. Additionally, PrivyTRAC improves the accuracy of infection risk estimation and consequently contact tracing effectiveness. As public health agencies are struggling to meet the expected demands of qualified personnel for traditional contact tracing measures, PrivyTRAC can be an effective tool to fill the resource gap.

The proposed capability also provides valuable actionable information for authorities to better allocate resources and plan follow-on actions. First, based on the estimated infection risks and visitors/foot traffics, authorities can identify and prioritize areas that require disinfection. Second, decision makers can pre-position test kits and allocate other health resources according to locations, populations and severity of the virus exposure. Similarly, when a person is potentially exposed to the virus, the user App can suggest local test locations that best balance test site workloads and convenience.

The system concept is applicable for COVID-19 as well as for other contagious diseases. By adjusting the disease vector, risk model and environmental effects, the system can be tailored to other infectious diseases in the future. The system architecture remains the same. The risk maps are tailored to different diseases in the same App environment.